# Interfacial nanostructure induced spin-reorientation transition in Ni∕Fe∕Ni∕W(110)

J.-S. Lee,[1] J. T. Sadowski,[2] H. Jang,[3] J.-H. Park,[3,4] J.-Y. Kim,[5] J. Hu,[6] R. Wu,[6] and C.-C. Kao[1]

[1]*Stanford Synchrotron Radiation Lightsource, SLAC National Accelerator Laboratory, Menlo Park, California 94025, USA*
[2]*Center for Functional Nanomaterials, Brookhaven National Laboratory, Upton, New York 11973, USA*
[3]*c–CCMR and Department of Physics, Pohang University of Science and Technology, Pohang 790-784, S. Korea*
[4]*Division of Advanced Materials Science, Pohang University of Science and Technology, Pohang 790-784, S. Korea*
[5]*Pohang Accelerator Laboratory, Pohang University of Science and Technology, Pohang 790-784, S. Korea*
[6]*Department of Physics & Astronomy, University of California Irvine, California 92697, USA*

We investigated the mechanism of the spin-reorientation transition (SRT) in the Ni/Fe/Ni/W(110) system using *in situ* low-energy electron microscopy, x-ray magnetic circular dichroism measurements, and first principles electronic structure calculations. We discovered that the growth of Fe on a flat Ni film on a W (110) crystal resulted in the formation of nanosized particles, instead of a uniform monolayer of Fe as commonly assumed. This interfacial nanostructure leads to a change of the system's dimensionality from two-dimensional- to three dimensional-like, which simultaneously weakens the dipolar interaction and enhances the spin-orbit coupling in the system and drives the observed SRT.

## I. INTRODUCTION

Ultrathin magnetic structures are one of the best examples of nanoscience and technology. The basic magnetic properties of the individual constituents in the system, such as magnetic moment and anisotropy, and the interactions among them are extremely sensitive to their atomic structure, size, and dimensionality. This extraordinary sensitivity has led to the discovery of many new physical phenomena and device concepts in magnetic storage over the past two decades.[1–4] However, the origin of this sensitivity is still not fully understood and presents one of the major challenges in nanomagnetism today. As an example, one of the most actively researched areas in the field of ultrathin magnetic structures is to understand and control the magnetic easy axis in ultrathin magnetic structures with the aim to meet the needs of high-density magnetic storage and magneto-optics recording. This effort has led to the development of many novel ultrathin magnetic film structures.[5,6] On the other hand, our understanding of magnetic anisotropy in these systems is still incomplete, in particular with regards to the connection between complex interfacial atomic structures and macroscopic magnetic behavior, which is due to the difficulty in characterizing these interfaces.

In this work, we focus on the physics of the spin-reorientation transition (SRT) behavior discovered in the Ni(1 ML)-Fe(1 ML)-Ni(8 ML) system; where ML denotes a monolayer.[7–9] Specifically, at room temperature, the easy axis of the entire system changes from in-plane (8ML Ni) → out-of-plane (1ML Fe/ 8ML Ni) → in-plane (1ML Ni/ 1ML Fe/ 8ML Ni) with successive deposition of additional

MLs of Fe and Ni. This observation has attracted considerable interest because it provides a new controlling parameter to engineer magnetic anisotropy in magnetic nanostructures.[8–10] More importantly, it challenges our understanding of magnetic anisotropy in ultrathin magnetic structures.

The observed exotic SRT is attributed to the interface term in the magnetocrystalline anisotropy (MCA) since Meyerheim *et al.*[10] recently demonstrated with *in situ* surface x-ray scattering that the change in the lattice strain of an entire Ni/Fe/Ni/W(110) film is insufficient to generate the observed SRT. Within the current theoretical framework, SRT in ultrathin magnetic films is understood in terms of competition between bulk and interface components, and the effect of lattice strain through magnetoelastic (ME) coupling in the phenomenological model of total magnetic anisotropy energy (i.e., effective magnetic anisotropy $K_{eff}$):[11,12]

$$K_{eff}t = 2(K_I + B_I e + D_I e^2) + (B_B e + D_B e^2)t - 2\pi M^2 t.$$

Here, $B_{I(B)}$ is the first-order interface (bulk) ME coupling constant and $D_{I(B)}$ is the second-order interface (bulk) ME coupling constant. These can be derived from reported structural parameters of Ni[13–15] while a variation of the in-plane strain ($e$) is negligible in the Ni-Fe-Ni/W system.[10] Hence, an adjustable parameter for this SRT is only the interface MCA, $K_I$—the reported $K_I$ for Ni(111) film is −0.22 erg/cm$^2$.[14,15] However, the following analysis shows that one has to assume unphysically large enhancement of spin-orbit coupling energy after the deposition of Fe layer to account for the observed SRT.

Figure 1(a) shows the variation of $K_{eff}t$ as a function of $K_I$ with the fixed thickness of Ni ($t = 8$ ML). For thick Ni film case, the estimated $K_{eff}$ was to be estimated about −0.762 erg/cm$^2$, leading to the in-plane easy axis. Moreover, we simulated the spatial distribution of this anisotropy energy as shown in Fig. 1(b). The minimized surface area is in-plane when $K_I = -0.22$ erg/cm$^2$. Even with $K_I = 0$, both $K_{eff}$ and the shape of the magnetic surface still point to an in-plane direction. Considering the SRT after deposition of 1 ML of Fe and a change in coercivity,[10] the total anisotropy energy must vary ~40%, while the sign changes from negative to positive, resulting in $K_{eff}t = +0.418$ erg/cm$^2$. This leads to $K_I = +0.96$ erg/cm$^2$ and the minimum surface area pointing in the out-of-plane direction. However, this change in $K_I$ is unphysical because the corresponding enhancement of the spin-orbit coupling energy ($E$soc)[11–13] after Fe growth would have to be ~500% — a value that is far too large.

In the following, we present a combined experimental and theoretical study of the spin-reorientation transition in Ni/Fe/Ni/W(110). The *in situ* low-energy electron microscopy (LEEM) measurements unambiguously revealed that the growth of the Fe ML does not form a wetting layer, but rather comprises a layer of nanoparticles. The x-ray magnetic circular dichroism (XMCD) results demonstrate a strong enhancement of SOC along the out-of-plane direction at the Fe-Ni interface due to the symmetry-breaking boundary caused by the nanoparticles. Therefore, after the growth of the Fe ML, we can consider the magnetic anisotropy as having undergone a dimensional crossover; a two-dimensional- (2D-) →three-dimensional- (3D-) like system that entails a reduced dipolar interaction, thereby significantly contributing to the interface MCA of the entire system. Furthermore, our state of the art electronic structure calculations for these systems agree well with the experimental findings.

## II. SAMPLES AND EXPERIMENTS

The single crystalline W(110) substrate was cleaned through cycles of oxidation at around 1500 °C and flash heating at 2000 °C. Subsequently, epitaxial films of Ni(8 ML), Fe(1 ML)/Ni(8 ML), and Ni(1 ML)/Fe(1 ML)/Ni(8 ML) were deposited on the W(110) substrate at $T$ = 300 K, maintaining the background pressure at ∼2 × 10−10 Torr. Low-energy electron diffraction patterns confirmed the epitaxy of the films. The XMCD was measured with 95% circularly polarized incident light at the elliptically polarized undulator beamline 2A at the Pohang Light Source; spectra were collected in the total electron yield mode at 300 K. Dichroic signals via $2p \rightarrow 3d$ dipole transitions represent the parallel ($\rho+$) and antiparallel alignment ($\rho-$) of the magnetization direction with respect to the photon helicity. At each data point, a 0.5-T pulse magnet switched along the easy-axis–in-plane case (30°) and out-of-plane (90°) incident angles. Figures 2(a) and 2(b) show the dichroic signals ($\rho+$ and $\rho-$) on the thick Ni and Fe films, respectively. The $\rho+$ and $\rho-$ spectra resulting from $2p \rightarrow 3d$ dipole transitions are divided roughly into $L_3$ ($2p_{3/2}$) and $L_2$ ($2p_{1/2}$) regions.

## III. RESULTS AND DISCUSSION

To explore microscopically the magnetic influence of the Fe ML, we collected XMCD measurements at the Ni and Fe $L_{2,3}$ edges to obtain quantitative information about the element-specific spin and orbital magnetic moments.[16] Figures 2(c) and 2(d) show the XMCD($\Delta\rho$) and its integration for each film. The $\Sigma(\Delta\rho)$ over the entire $L_{2,3}$ region is proportional to the orbital magnetic moment. Using the sum rule,[16] we estimated the spin ($m_s$) and orbital ($m_o$) magnetic moments of Ni or Fe ions for each film. The total Ni moment ($m_s + m_o$) is similar to that of bulk Ni (∼0.7$\mu$B/Ni). For Fe/W, the Fe spin moment is smaller than that in a thick Fe film, while all $m_o$ values are similar to the bulk value. Since the Fe layer is very thin, the alignment of the spin shows characteristics of an ultrathin film (i.e., the temperature effect of magnetization). In this fashion, the spin fluctuates somewhat at the room temperature.[17] On the other hand, relatively all orbits are fully polarized, thereby increasing the moment ratio ($m_o/m_s$); we represent this enhancement from the bulk ratio as $\delta_{m_O/m_S}|^{ion} = [(m_o/m_s)^{ion} - (m_o/m_s)^{bulk}]/(m_o/m_s)^{bulk}$. Interestingly, after depositing Fe(1 ML), the change in the $\delta_{m_O/m_S}$ is remarkable. We recorded a ~220% enhancement for $\delta_{m_O/m_S}|^{Fe}$ while, for Ni, the increase is about ~25%. This signifies that the strongly increased Fe $E_{soc}$ after the growth of Fe (1 ML) affects Ni $E_{soc}$ along the out-of-plane direction. Furthermore, we note a possible intermixing at the Fe-Ni interface. Since the ground state of bulk Ni can be represented as multiplet configuration states,[18] the absorption and its XMCD spectra additionally exhibit the so-called eV and 4-eV satellites above the higher-energy region of the $L_{2,3}$ main features, respectively. In particular, this satellite feature is regarded as a good indicator of the alloy or intermixing effect.[19,20] In these systems, the spectral shapes are the same as for thick films (i.e., there is no change in the spectral shapes). In this fashion, we could rule out the possibility of intermixing as an origin of the enhancement of the orbital moment.

Based on these XMCD results, the enhanced SOC after Fe growth is around 120%. Since the structural modifications (i.e., strain) after Fe growth are negligible,[10] this SOC enhancement seemingly is associated only with the interface-MCA effect. Although the change in $K_I$ via $E_{soc}$ is pronounced, however, this heightening is clearly insufficient to account for the observed SRT behavior (i.e., ~500%), as we estimated. Accordingly, there must be an additional mechanism operating. One possible scenario proposes a major modification in the dipolar interactions in the whole system associated with the deposition of Fe. For example, if there is a 2D → 3D (or 1D) transition, the contribution of dipolar interaction energy ($E_{dip}$) to the total magnetic anisotropic energy might be lowered dramatically.[5] Since the primary effect of the in-plane magnetic anisotropy in the 2D system reflects $E_{dip}$, then $K_{eff}t$ also may be significantly affected should there be a dimensional alteration. To examine this possibility, we carried out *in situ* LEEM observations of each stage of the growth of the Ni/Fe/Ni/W(110) structure. LEEM measurements were undertaken with the Elmitec system (LEEM III) at beamline U5UA at the National Synchrotron Light Source (NSLS). LEEM is well suited for this task because it supports *in situ* monitoring of the change of surface morphology with a lateral resolution of a few nanometers and a single-atomic-layer resolution in the direction normal to the substrate surface.

Figure 3 shows series of the LEEM images obtained from consecutive layers in the Ni/Fe/Ni/W(110) structure (top), and their corresponding schematic topological map and spin configurations (bottom). We note that the images were taken at 1.0 eV. The nonmagnetic W(110) substrate is clean, as evidenced by the array of monolayer-high steps visible in Fig. 3(a). Consecutively, to assure a flat, well-ordered surface of Ni(8 ML), we followed a well-known procedure:[21] (i) we deposited only Ni(1ML) on W(110); (ii) the film was annealed to ~900 K for several minutes to promote formation of a well-ordered *c*-(1 × 7) structure, as monitored by the *in situ* LEED pattern; and (iii) we deposited the remaining 7ML of Ni, obtaining a flat surface [Fig. 3(b)]. At that time, the magnetic easy axis was in-plane, and the dominant magnetic behavior in the flat Ni(8 ML) film was along the in-plane direction via $E_{dip}$, including the negative $K_I$.

After depositing Fe(1 ML), the LEEM image changed dramatically. Figure 3(c) shows the many sizable spots that became evident. The mosaic-like contrast features indicate the presence of nanosized Fe particles (Fe$^{Par}$), signifying that there was a 2D- → 3D-like change in dimension of this system. Since the mosaic-like LEEM contrast may also originate from the misoriented grains forming a continuous film, we checked the LEED measurement, showing 1 × 1 hexagonal pattern, and ruled out the possibility of any domains or grains. The shapes of the particles, including their size, are somewhat random (represented by the distributions of circles and squares in the figure, and the average size ∼100 nm). These features point to a break in symmetry at the boundary between Ni and Fe$^{Par}$, generating the pronounced $E_{soc}$ between them along the out-of-plane direction that reinforces the interface MCA of the entire system. Concomitantly, the $E_{dip}$ along the in-plane direction weakens because of the gap between Fe$^{Par}$ neighbors. Consequently, the combination of both magnetic energies is responsible for the SRT (i.e., in-plane → out-of-plane) after the deposition of a ML of Fe. Furthermore, the small spin moment on the Fe(1 ML) in the XMCD findings might reflect the disordered spin state around the particle surfaces.

LEEM images [Fig. 3(d)] taken from the Ni(1 ML)/Fe/Ni/W film shows that the surface has become smooth again. This observation indicates that adding the Ni(1 ML) fills the gaps between neighboring Fe particles, with only small deposits of Ni on top of Fe$^{Par}$, entailing a reverse 3D-like → 2D transition. Accordingly, $E^{dip}$ becomes dominant again, and the magnetic easy axis naturally changes from the out-of-plane direction into the in-plane direction. We also note that further investigations using a kind of microsimulation may support the size effect of the particles.

Aside from the change is shape anisotropy as discussed above, an important contribution to SRT is the change of interfacial bonding. Finally, we performed first-principles calculations with the all-electron full-potential linearized augmented plane wave (FLAPW) method.[22] To shed light on this aspect, magnetocrystalline anisotropy energies ($E$MCA) of Ni(111), Fe/Ni(111), and Ni/Fe/Ni(111) films were calculated with the torque approach,[23] and results are presented in Fig. 4. A flat Fe ML on Ni(111) depicted in model *A* has the highest total energy, 0.62 eV. This means that the flat Fe formation is very unstable on top of Ni(111), which is consistent with LEEM results. The value of uniaxial $E_{MCA}$ (i.e., the energy difference for magnetization pointing along the *z* and *x* axes) is −0.67 erg/cm$^2$ for a 13 ML Ni(111), in excellent agreement with our experiment. Note that the formation of rough morphology in Fig. 3(c) is schematically represented by a little bit of Fe diffusion. For Fe/Ni(111), both *B* and *C* models have positive $E$MCA, leading to the SRT as we observed. Hence, the effect of the rough morphology of Fe on the SRT is twofold: reducing the shape anisotropy through morphology change and also promoting $E_{MCA}$. In contrast, we obtained negative $E_{MCA}$ for models *D*, *E*, and *F*, indicating that the magnetization direction of Ni/Fe/Ni(111) film becomes in-plane again with the presence of a Ni adlayer. It appears that positive $E_{MCA}$ results from Fe-Fe and Fe-Ni bonding in the two outermost layers. The interesting aspect of this calculation is that such twofold effects are coherent for both Fe/Ni and Ni/Fe/Ni. To show if one can manipulate $E_{MCA}$, we also give curves of $E_{MCA}$ versus the change in the Fermi level, calculated using a rigid-band approximation. Overall, these curves are very smooth in a broad energy range, indicating the quality of our theoretical results. It is striking that a change of $E_{MCA}$ sign is variable if the Fermi level is shifted by ±0.06 eV. This indicates a good opportunity to tune the magnetization of Fe/Ni(111) with electric field.

## IV. SUMMARY

In summary, we verified that the presence of Fe nanoparticles determine and control the stability of the magnetic anisotropy in an Ni/Fe/Ni system. The growth and formation of the Fe nanoparticles modulates the dimensions of the thin-film system. Due to the subsequent induced symmetry breaking at the boundary of the particle layer and the gaps between the particles, both spin-orbit coupling and dipolar interactions are strongly modified, resulting in the spin-reorientation transitions in Ni/Fe/Ni/W system. Our results demonstrate a role of the magnetic properties of these finite-size particles. Further investigation, such as magnetic imaging, is underway to investigate the correlation of the magnetic domain motions.

## ACKNOWLEDGMENTS

NSLS and research at the Center for Functional Nanomaterials, BNL, are supported by the US DOE, Office of Science, Office of Basic Energy Sciences, under Contract DE-AC02-98CH10886. POSTECH is supported by the National Creative Initiative Center for *c*−CCMR (2009-0081576), WCU program (R31-2008-000-10059-0), and Leading Foreign Research Institute Recruitment program (2010-00471) through NRF funded by MEST. PAL is supported by POSTECH and MOST. Work at the University of California, Irvine is supported by DOE grant DE-FG02-05ER46237. Calculations were performed on parallel computers at NERSC.

**[Figures]**

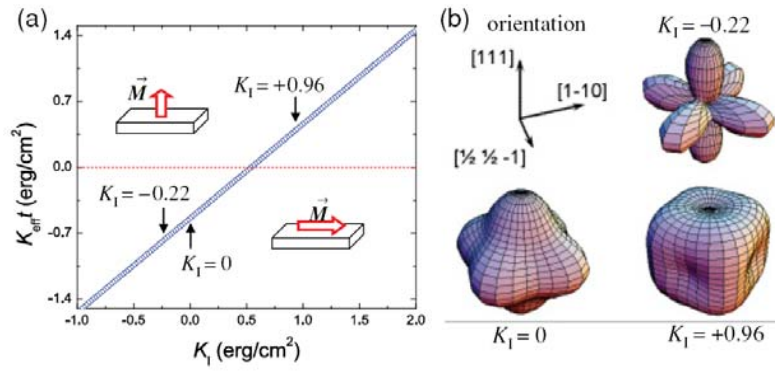

FIG. 1. (Color online) (a) Simulation of the effective magnetic anisotropy ($K_{eff}t$) versus the interface magnetic anisotropy ($K_I$) on Ni(8 ML)/W(110). (b) Simulation of magnetic anisotropy energy surface for $K_I = -0.22$, $= 0$, and $= +0.96$ erg/cm$^2$.

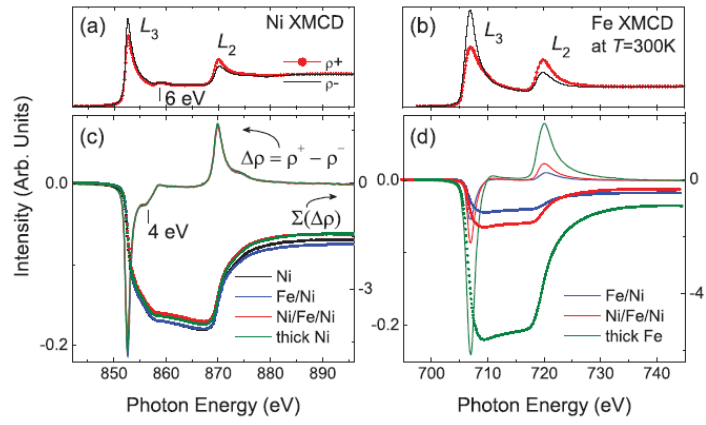

FIG. 2. (Color online) Dichroism ($\rho+$ and $\rho-$) of thick Ni (a) and Fe (b) films. (c), (d) $\Delta\rho = \rho+ - \rho-$ and its integration, $\Sigma(\Delta\rho)$, for Ni (black), Fe/Ni (blue), and Ni/Fe/Ni (red) films. The vertical bars in Ni spectra denote 4- and 6-eV satellites.

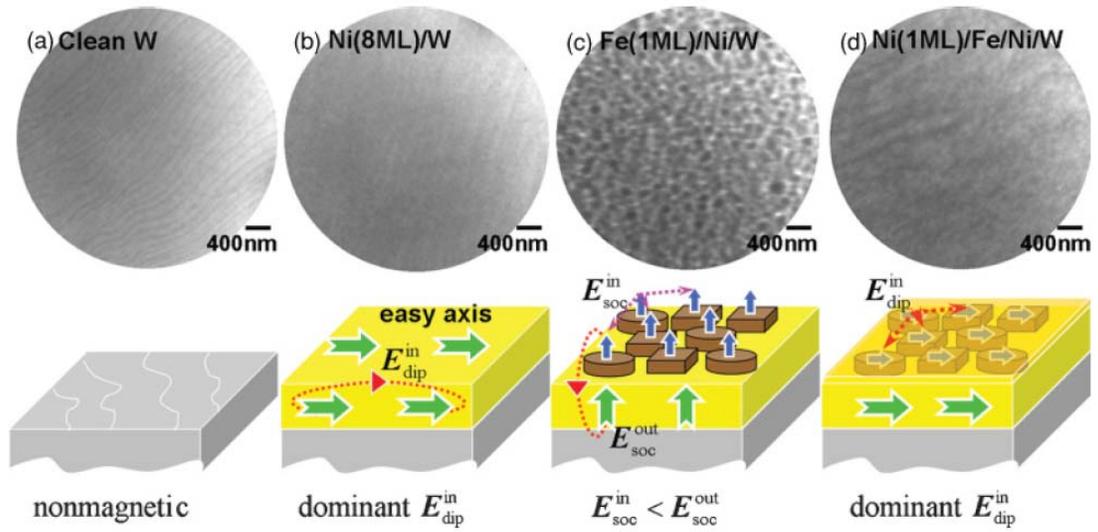

FIG. 3. (Color online) (top) Series of LEEM images obtained from consecutive layers in the Ni/Fe/Ni/W(110) structure, and (bottom) the scheme of magnetic interactions via spin-orbit coupling ($E_{soc}$) or dipolar interactions ($E_{dip}$) on each film: (a) W(110), (b) Ni(8 ML)/W, (c) Fe(1ML)/Ni(8 ML)/W, and (d) Ni(1 ML)/Fe(1 ML)/Ni(8 ML)/W. Green and blue arrows denote the spin direction in the film. The dotted line represents the magnetic interaction.

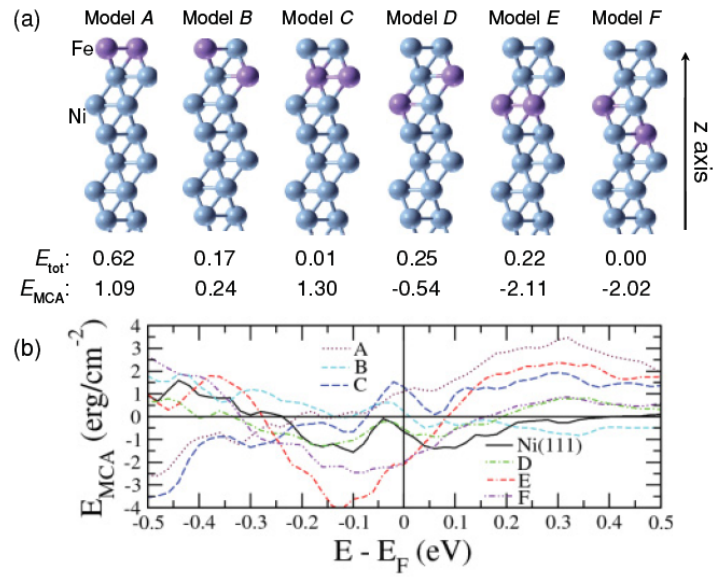

FIG. 4. (Color online) (a) Surface models for Fe/Ni(111), with light blue and purple balls for Ni and Fe atoms, respectively. The total energies, $E_{tot}$ (in eV) are measured from the reference system, model $F$. The MCA energies (in erg/cm$^2$) are for the energy differences between the $z$ and $x$ axis. (b) First-principles calculations with the all electron FLAPW method on the models and bulk Ni(111) reference.